\documentclass[aps,prl,preprint,superscriptaddress,longbibliography]{revtex4-1}
\usepackage{graphicx}
\usepackage{float}
\usepackage{amsmath}
\usepackage{color}

\begin{document}


\title{Electron Doping a Kagom\'{e} Spin Liquid}



\author{Z.A. Kelly}
\affiliation{Department of Chemistry, The Johns Hopkins University, Baltimore, MD 21218}
\affiliation{Institute for Quantum Matter, Department of Physics and Astronomy, The Johns Hopkins University, Baltimore, MD 21218}
\author{M.J. Gallagher}
\affiliation{Department of Chemistry, The Johns Hopkins University, Baltimore, MD 21218}
\author{T.M. McQueen}
\email{mcqueen@jhu.edu}
\affiliation{Department of Chemistry, The Johns Hopkins University, Baltimore, MD 21218}
\affiliation{Institute for Quantum Matter, Department of Physics and Astronomy, The Johns Hopkins University, Baltimore, MD 21218}
\affiliation{Department of Materials Science and Engineering, The Johns Hopkins University, Baltimore, MD 21218}%


\date{\today}

\begin{abstract}
Herbertsmithite, ZnCu$_3$(OH)$_6$Cl$_2$, is a two dimensional kagom\'{e} lattice realization of a spin liquid, with evidence for fractionalized excitations and a gapped ground state. Such a quantum spin liquid has been proposed to underlie high temperature superconductivity and is predicted to produce a wealth of new states, including a Dirac metal at $1/3$rd electron doping. Here we report the topochemical synthesis of electron-doped ZnLi$_x$Cu$_3$(OH)$_6$Cl$_2$ from $x$ = 0 to $x$ = 1.8 ($3/5$th per Cu$^{2+}$).  Contrary to expectations, no metallicity or superconductivity is induced. Instead, we find a systematic suppression of magnetic behavior across the phase diagram. Our results demonstrate that significant theoretical work is needed to understand and predict the role of doping in magnetically frustrated narrow band insulators, particularly the interplay between local structural disorder and tendency toward electron localization, and pave the way for future studies of doped spin liquids.
\end{abstract}


\maketitle

\begin{figure}[H]
\includegraphics[width=\linewidth]{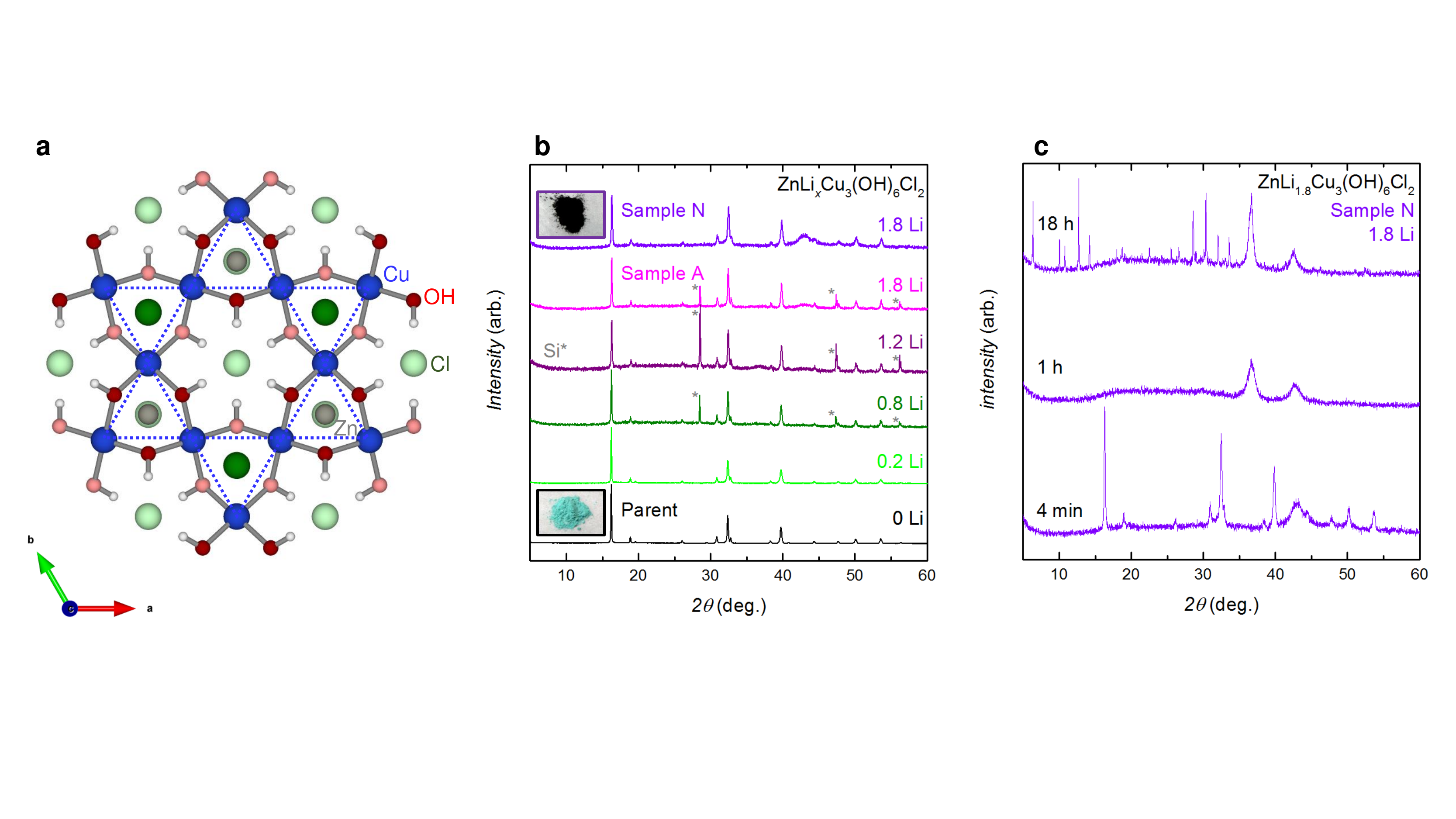}
\caption{\label{Figure 1} Doped Herbertsmithite structure. (a) a top-down (along c-axis) representation of the parent Herbertsmithite copper kagom\'{e} layer (blue dotted line) with Cu (blue) and O (red), H (white), Zn (gray) and Cl (green) between the kagom\'{e} layers. The dark and light atoms are located above and below the kagom\'{e} plane respectively. (b) The X-ray powder diffraction (XRPD) patterns of the complete series. The gray asterisks represent the presence of Si (internal standard). An image of the blue-green parent material is shown in the lower left corner while a picture of the black doped sample N is shown in the upper left. All doped samples are also black. (c) XRPD data demonstrates the instability of one of the maximally doped samples, sample N, in air ($x = 1.8$) as it decomposes in hours into several other phases.}
\end{figure}

For decades, the resonance valance bond (RVB), or quantum spin-liquid, state has been theorized to be an intricate part of the mechanism for high temperature superconductivity\cite{Anderson1973,Anderson1987}. One geometrically frustrated system, Herbertsmithite (Fig.\ref{Figure 1}(a)), is considered an ideal spin two dimensional liquid candidate due to its perfectly ordered kagom\'{e} lattice of $S = 1/2$ copper ions, antiferromagnetic interactions with $J\approx-200$ K, strong evidence for fractional spin excitations by neutron scattering, and, most recently, convincing indications of a gapped spin-liquid ground state by oxygen-17 NMR\cite{Fu2015,Balents2010,shores2005,Freedman2010,Chu2010,Han2012}. All of these factors suggest Herbertsmithite is the realization of a quantum spin liquid. Recent predictions expanded upon Anderson’s theory in DFT calculations of electron doped Herbertsmithite, M$_x$Zn$_{1-x}$Cu$_3$(OH)$_6$Cl$_2$, where Ga$^{3+}$ or other aliovalent metals replace zinc\cite{Mazin2014,Guterding2015}. A trivalent substitution introduces electrons into the material, raising the Fermi level to the Dirac points at $x = 1$, and giving rise to a rich phase diagram spanning from a frustrated RVB spin liquid ($x = 0$) to a strongly correlated Dirac metal ($x = 1$) with possible Mott-Hubbard metal-insulator transitions, charge ordering, ferromagnetism, or superconducting states. 

It is challenging to synthesize electron doped Herbertsmithite directly as Cu$^{1+}$ will not assume the same distorted octahedral site on the kagom\'{e} lattice as Cu$^{2+}$ under thermodynamic conditions, and copper(I) hydroxide is thermodynamically unstable towards disproportionation and evolution of hydrogen gas. By using low temperature topochemical techniques, this problem is circumvented by producing a kinetically meta-stable phase\cite{Hayward2002,Tsujimoto2007,Rosseinsky1992,Neilson2012}. Here we use intercalation of lithium to produce electron doped Herbertsmithite, ZnLi$_x$Cu$_3$(OH)$_6$Cl$_2$ with $0 \leq x \leq 1.8$.
%

Laboratory X-ray powder diffraction (XRPD), Fig.\ref{Figure 1}(b), shows the underlying structure is maintained throughout the doped series. Lithium is not directly detected due to its small X-ray scattering intensity relative to copper and zinc. Any changes in the lattice parameters as a function of doping are small and are within the resolution of the Laboratory X-ray diffractometer (see SI). During Rietveld analysis, CuO and Cu$_2$O were tested and are absent from the air-free samples by both XRPD and neutron diffraction. Unlike the air stable parent, the doped samples decomposed readily in air, Fig.\ref{Figure 1}(c), with the most heavily doped samples completely decomposing within hours. This rapid and total decomposition is in agreement with the formation of a reduced copper (Cu$^{1+}$) hydroxide in the bulk that is prone to decomposition in moisture. The color change from blue to black is also in agreement. As soon as there are any Cu$^{1+}$ ions present, there is another possible optical absorption mode: intervalence charge transfer (i.e. Cu$^{2+}$ + Cu$^{1+}$ $\rightarrow$ Cu$^{1+}$ + Cu$^{2+}$), or, put another way, a transition from an “impurity band” in the gap to the conduction band. Such absorption modes are common in mixed valent systems, such as the Cu$^{1+}$ - Cu$^{2+}$ mixed valence (N$_2$H$_5$)$_2$Cu$_3$Cl$_6$\cite{Scott1991}.

To determine the position of Li within the structure, we carried out neutron powder diffraction of the undoped and maximally Li-doped specimens using the high flux NOMAD diffractometer at the Spallation Neutron Source, Oak Ridge National Laboratory (see SI). Rietveld analysis reveals that the previously reported structure accurately models the data of the doped specimens, with the exception of the presence of a pocket of negative scattering in a tetrahedral hole formed by three (OH$^-$) and one Cl$^-$ group, located above and below the copper triangles in the kagom\'{e} layer. This is consistent with the presence of Li, which has a negative scattering factor. Although the site is physically small for a Li ion, the connectivity is consistent with a favorable tetrahedral bonding environment for Li. The XRPD studies are also consistent with this model. There are systematic changes in the O-Cu-Cl bond angle and the O-Cu, Cl-Cu, and O-O bond lengths (see SI). As the doping increased, the oxygen atoms move away from the Cu kagom\'{e} lattice and spread from one another. In concert, the Cl atom moves away from the kagom\'{e} lattice along the c-axis. These combined movements create more space in the Cl-(OH)$_3$ tetrahedral hole. Further, a similar geometry is found in CuMg$_2$Li$_{0.31}$\cite{Braga2007}, and a stable Rietveld refinement is obtained for the maximally doped sample N, when including Li in that site, with the occupancy refining to $\sim$0.9 ($x = 1.8(3)$ per formula unit, see SI). This structure puts the Li ion in close proximity to the Cl atom and appears to form a neutral LiCl dimer along the c-axis with a bond distance of $\sim$1.4 \AA. Such a dimer is consistent with our attempts to intercalate the larger K$^+$ ion, which resulted instead in the formation of KCl. Future work is needed to determine if this model is an accurate description of the local atomic structure.

\begin{figure}[H]
	\includegraphics[width=\linewidth]{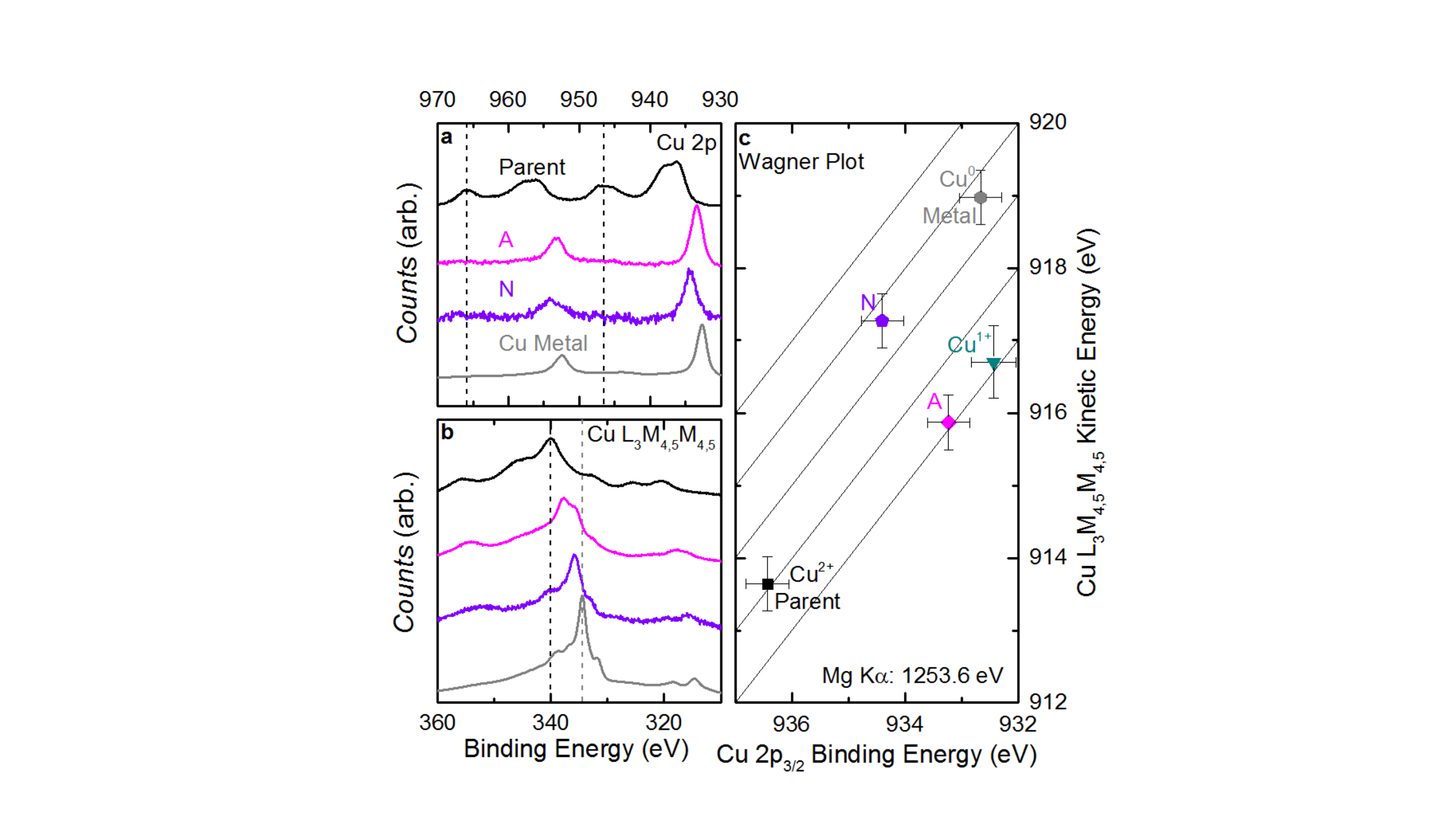}
	\caption{\label{Figure 2} X-ray Photoelectron Spectroscopy (XPS). (a) Cu 2p XP spectra of parent Herbertsmithite (black), Sample A (magenta), Sample N (violet), and Cu metal (gray). The black dashed line indicates locations of satellite peaks in the parent, characteristic of Cu$^{2+}$, which are significantly reduced in the doped samples. (b) The X-ray generated Auger Cu L$_3$M$_{4,5}$M$_{4,5}$ spectra of the same four samples. The black and the gray dotted lines represent the location of the greatest intensity peak for the parent and the copper metal respectively. The peak shape and binding energy of the doped samples varies significantly from both the parent and the copper metal. (c) A Wagner plot shows the relative chemical shift of the four samples and Cu$^{1+}$ in Cu(I)$_2$O (lit. teal)\cite{NIST2012} by plotting the kinetic energy from the Cu  L$_3$M$_{4,5}$M$_{4,5}$ peak on the y-axis and the binding energy from the Cu 2p$_{3/2}$ peak on the x-axis. The chemical shift is sensitive to the polarizability of the chemical environment} 
\end{figure}

X-Ray Photoelectron Spectroscopy (XPS) provides a direct probe of the chemical environment of copper and was carried out on the parent and two maximally doped specimens, A and N. The results are consistent with the reduction of Cu$^{2+}$ to Cu$^{1+}$. Firstly, the four peaks in the parent Cu 2p envelope, Fig.\ref{Figure 2}(a), are indicative of the two final states in divalent copper, (i) the  3d$^{10}$L$^{-1}$ due to an exiting photoelectron leaving a core hole causing a charge transfer process between the surrounding ligands and Cu d shell and (ii) the 3d$^{9}$L satellite. In the doped samples, this satellite is greatly reduced due to the filled 3d shell in Cu$^{1+}$ preventing this loss transition from occurring\cite{McIntyre1975,VanderHeide2011}. If it were purely Robin-Day Class 1 mixed valance (pure Cu$^{1+}$ and Cu$^{2+}$ sites with no interactions of ground or excited states), we would expect a mixed XPS Signal of Cu$^{1+}$ and Cu$^{2+}$ with an approximate 2:1 ratio. In this case, however, there must be interactions between neighboring Cu$^{1+}$ and Cu$^{2+}$, given the shared hydroxyl bridge, through which we know (from the parent) that adjacent Cu ions interact\cite{Robin1967,Day2008,Rupp1976}. The result is a suppression of the Cu$^{2+}$ XPS satellites, even though resistance measurements show the charges must be localized. This model (which has discrete Cu$^{1+}$ and Cu$^{2+}$ ions, Robin-Day Class 2), would not only suppress the Cu$^{2+}$ satellites but also give rise to an optical intervalence charge transfer, which would explain the black color of the material upon even light doping.

Secondly, the photoelectron induced Auger Cu L$_3$M$_{4,5}$M$_{4,5}$ spectra, Fig.\ref{Figure 2}(b), of the maximally doped specimens are in between and distinct from the L$_3$M$_{4,5}$M$_{4,5}$ spectra of the Cu foil and the parent Herbertsmithite. Further, a Wagner plot analysis, Fig.\ref{Figure 2}(c), shows that the Li doped samples are in a distinctly different chemical environment than either the parent (fully Cu$^{2+}$) or Cu metal (fully Cu$^0$)\cite{Wagner1972}, consistent with the structure suggested by our neutron diffraction studies and indicative of the presence of Cu$^{1+}$\cite{McIntyre1975,VanderHeide2011}. Although information on copper oxidation states is lost in a depth profile analysis with ion sputtering, it can be used to determine the chemical composition\cite{panzner1985}. As expected from the topochemical synthesis method, a thin surface layer of Li and benzophenone starting material is detected; upon ion sputtering (up to 100 min), the ratio of Cu:Zn:Cl is in agreement with the expected parent Herbertsmithite phase, with Li located throughout (see SI). 

\begin{figure}[H]
	\includegraphics[width=\linewidth]{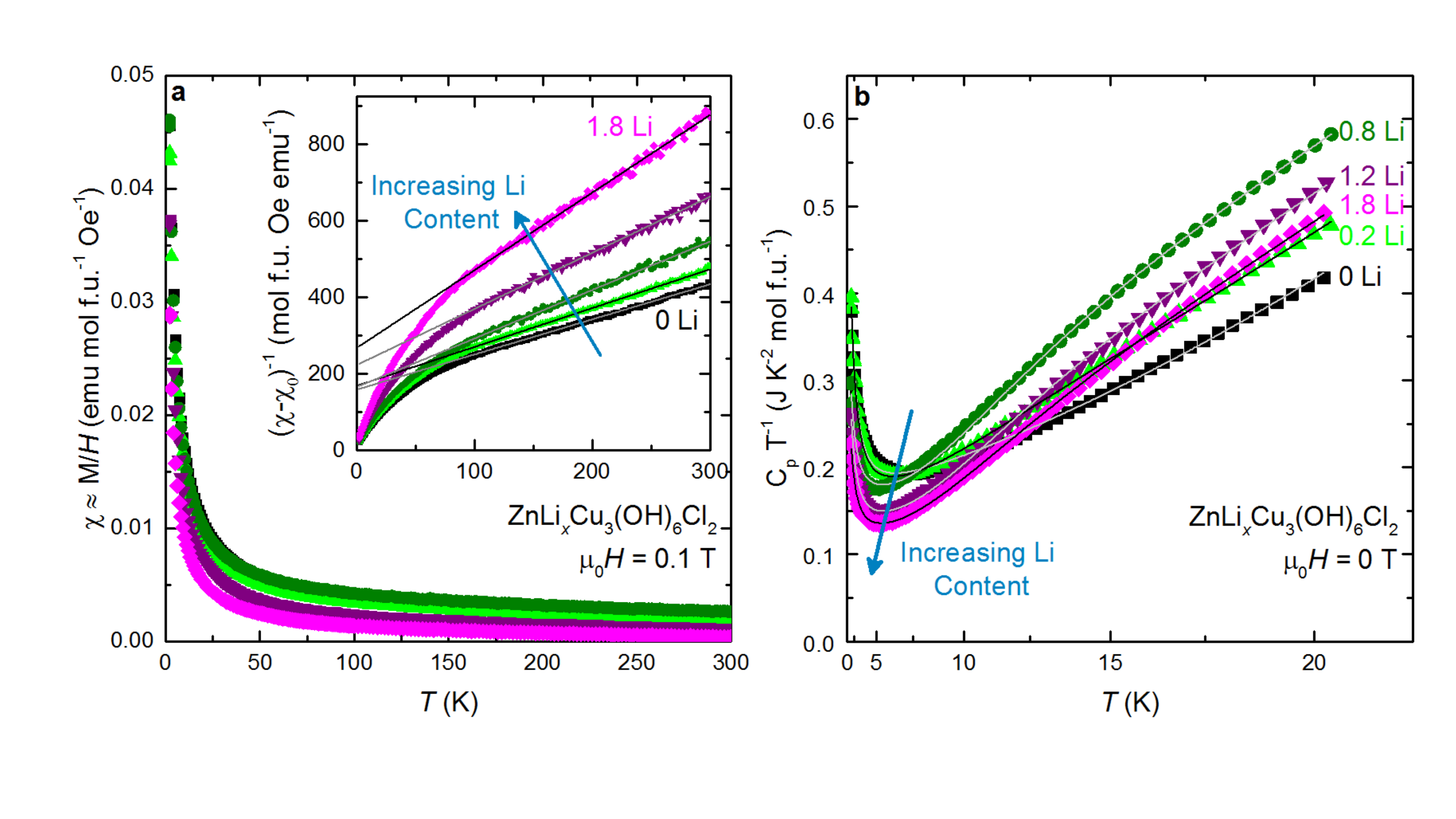}
	\caption{\label{Figure 3} Physical properties of ZnLi$_x$Cu$_3$(OH)$_6$Cl$_2$ series. (a) The magnetic susceptibility, $\chi\approx M/H$, as a function of temperature for the doped ZnLi$_x$Cu$_3$(OH)$_6$Cl$_2$ series. Black and gray lines are from high temperature Curie-Weiss analysis of the $\chi_0$-corrected inverse magnetic susceptibility (inset). All samples have paramagnetic behavior and a decrease in susceptibility is seen as Li content increases. (b) Heat capacity divided by temperature as a function of temperature under zero field from $T=$1.8-20 K. The low temperature region systematically decreases with increasing Li content across the series, while at higher temperatures the doped samples have increased entropy. Black and gray lines represent fits to the data.}
\end{figure}
\begin{figure}[H]
	\includegraphics[width=\linewidth]{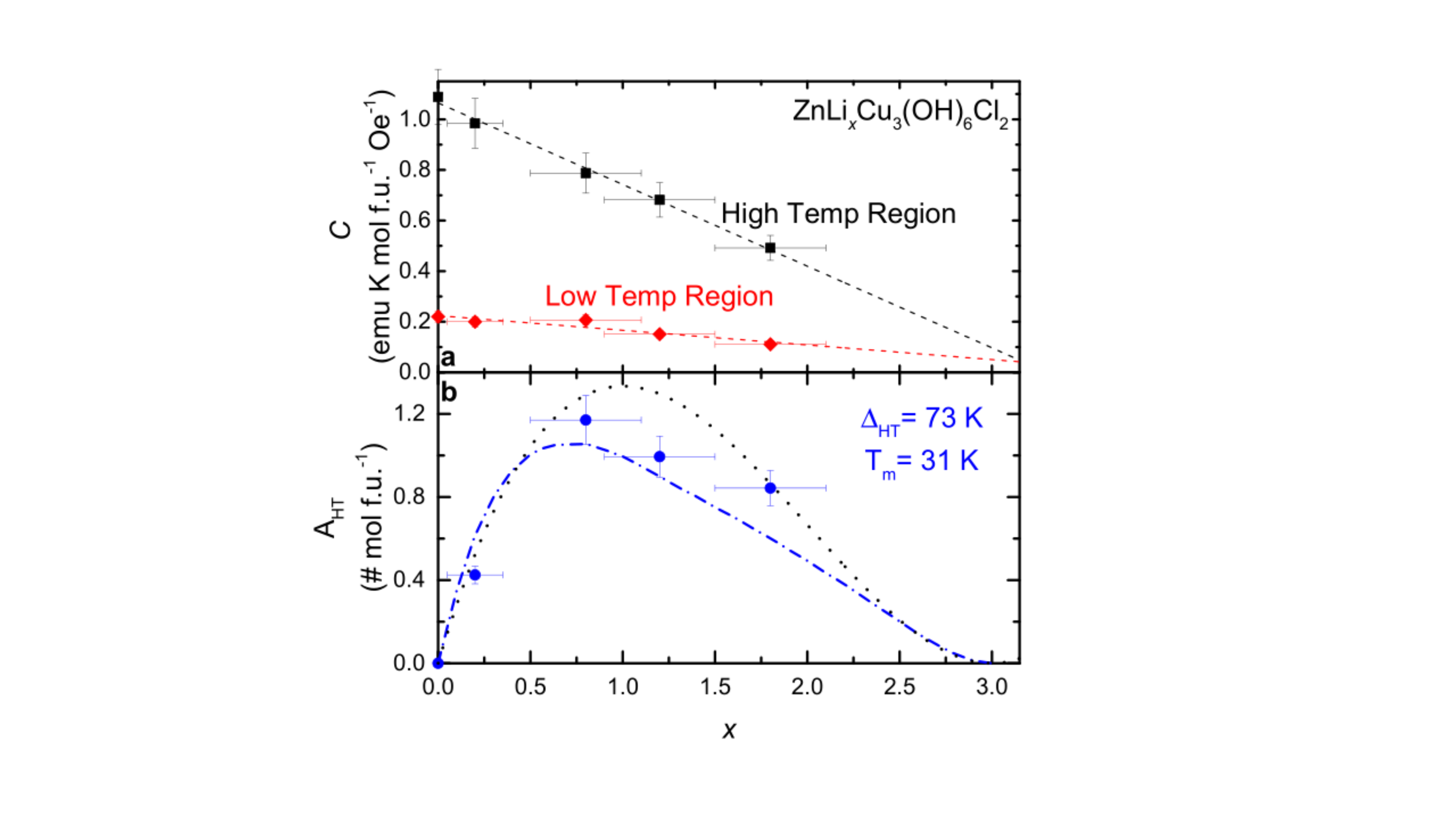}
	\caption{\label{Figure Combo} Magnetization and heat capacity fit parameters. (a) The extracted Curie constants, $C$, from the high (black squares) and low (red diamonds) temperature Curie-Weiss analysis of the ZnLi$_x$Cu$_3$(OH)$_6$Cl$_2$ series. The dashed lines are a guide to the eye that demonstrate a linear decrease. (b) The Schottky anomaly parameter, $A_{HT}$ (blue circles), from heat capacity fits to the doped Herbertsmithite series, which describes the feature in the high temperature heat capacity data. The blue dashed line and the black dotted line are two different models for singlet trapping in doped Herbertsmithite (see SI).}
\end{figure}
\begin{figure}[H]
	\includegraphics[width=\linewidth]{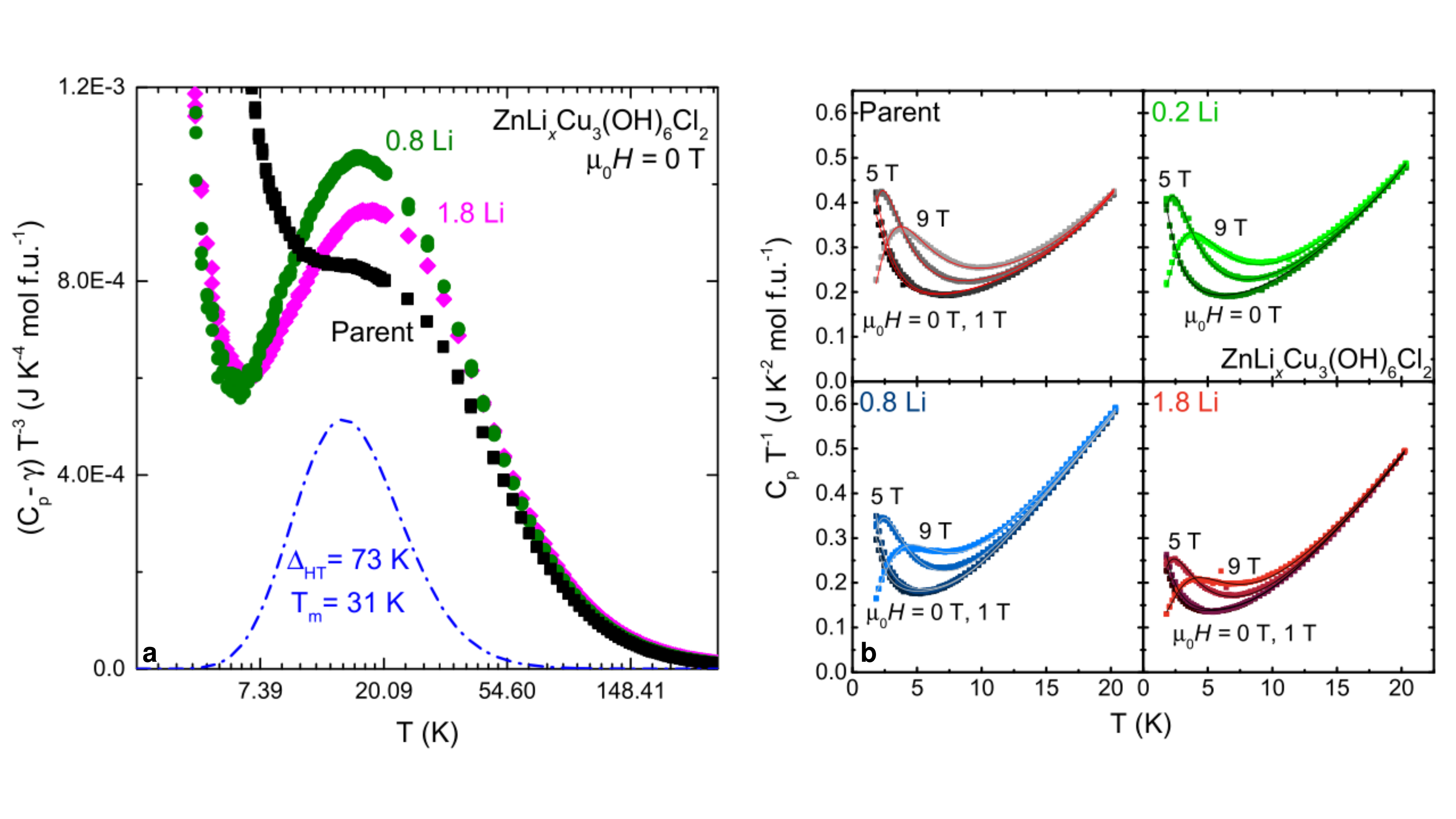}
	\caption{\label{HeatCap2} Detailed heat capacity analysis of ZnLi$_x$Cu$_3$(OH)$_6$Cl$_2$ series. (a) High temperature ($C_{p}-\gamma$) $T^{-3}$ Schottky analysis of the temperature range $T=$1.8-300 K for the parent ZnCu$_3$(OH)$_6$Cl$_2$ (black), ZnLi$_{0.8}$Cu$_3$(OH)$_6$Cl$_2$ (dark green), and ZnLi$_{0.8}$Cu$_3$(OH)$_6$Cl$_2$ (magenta). The dashed blue line is the high temperature Schottky anomaly of the extracted values from the zero field $T=$ 1.8-20 K fits. It does a good job of describing the Schottky anomaly we see in the doped samples in higher temperature region of the heat capacity. Debye modes (which are a constant at low temperature and fall off at higher temperatures) are not shown for clarity. (b) Simple model fits to field dependent heat capacity measurements. (top left) the Parent ZnCu$_3$(OH)$_6$Cl$_2$ (black), (top right) ZnLi$_{0.2}$Cu$_3$(OH)$_6$Cl$_2$ (green), (bottom left) ZnLi$_{0.8}$Cu$_3$(OH)$_6$Cl$_2$ (blue), and (bottom right) ZnLi$_{1.2}$Cu$_3$(OH)$_6$Cl$_2$ (red). The data sets go from a dark color at low fields to a lighter color at higher fields. The lines are fits as described in the text.}
\end{figure}
Despite the introduction of a substantial number of electrons, the material remains insulating: two probe room temperature resistance measurements on cold pressed pellets in a glovebox give a resistance $>$ 2 M$\Omega$ for the doped series. Fig.\ref{Figure 3}(a) shows the magnetic susceptibility, $\chi \approx M/H$, for the ZnLi$_x$Cu$_3$(OH)$_6$Cl$_2$ series. For $x = 0$, the inverse magnetic susceptibility is well-known to be linear at high temperatures and dominated by the kagom\'{e} network, with the signal at $T < 20$ K containing significant contributions from defect Cu$^{2+}$ ions on the Zn$^{2+}$ site between kagom\'{e} layers\cite{shores2005}. We thus performed fits to the Curie-Weiss law in the low temperature ($T$ = 1.8-15 K) and high temperature ($T$ = 100-300 K) regions to extract estimates of the number of spins arising from the intrinsic and excess Cu ions respectively as a function of $x$. The extracted Curie constants of both the low and high temperature regions decrease linearly with increasing doping level, Fig.\ref{Figure Combo}(a). This systematic decrease is consistent with the reduction of magnetic Cu$^{2+}$ ($S = 1/2$) to non-magnetic Cu$^{1+}$ ($S = 0$). With an x-intercept value of $x = 3.3(5)$, the high temperature extrapolation to zero is also consistent with the known stoichiometry of Herbertsmithite, Zn$_{0.85}$Cu$_{3.15}$(OH)$_6$Cl$_2$, where $x = 3.15$ would be necessary to convert all Cu$^{2+}$ to Cu$^{1+}$. All of the Weiss temperatures are negative, becoming less negative upon doping (see SI), in agreement with the expectation that the number of spins are reduced in the lattice. The low temperature extrapolation x-intercept value is $x=3.9(9)$; this is within error equal to that found from the high temperature extrapolation. Any subtle divergence between the high and low temperature x-intercept likely reflects a difference in reducibility of the kagom\'{e} compared to the interlayer Cu$^{2+}$ ions, since the high temperature paramagnetism includes both the kagom\'{e} and interlayer spins, whereas the latter is attributable only to the interlayer ‘defect’ spins. Given the placement of the Li ions near the kagom\'{e} layer, it is no surprise the kagom\'{e} layers are more greatly reduced than the interlayer sites. Further, the difference in local coordination (interlayer Cu in O$_{6}$ octahedron vs kagom\'{e} Cu in O$_{4}$Cl$_{2}$ octahedron), would result in a difference in redox potential for Cu$^{2+}$ + e$^-$ $\rightarrow$ Cu$^{1+}$ between the two sites, so reducing one should be slightly more favorable than reducing the other.

Fig.\ref{Figure 3}(b) shows the low temperature heat capacity. There are two regions of significant entropy change as a function of doping: at $T \approx$ 5 K, the heat capacity of the sample decreases with increasing Li content while at higher temperatures, there is an entropy gain at non-zero $x$. Qualitatively, the low temperature data can be explained by same mechanism as the magnetization, namely a reduction of the number of spins as Cu$^{2+}$ is converted to Cu$^{1+}$. To more quantitatively describe the changes, we parameterized the temperature-dependent data as a function of composition and applied magnetic field with the model:
\begin{equation}
\label{eqn 1}
C_p = \gamma T + \beta_3 T^3 + \beta_5 T^5 + A_{LT} f(\Delta_{LT},T) + A_{HT}f(\Delta_{HT},T)
\end{equation}
\begin{equation}
\label{eqn 2}
A f(\Delta,T) = AR(\Delta/T)^2\frac{e^{\Delta/T}}{(1+e^{\Delta/T})^2}
\end{equation}
The $\gamma T$ term captures the linear contribution to the specific heat from the spin liquid (either intrinsic or due to defect spins). The phonon contribution is described by the $\beta_3 T^3$ and $\beta_5 T^5$ terms\cite{Tari2003}. These phonon terms were calculated based on the field fit to the parent. The terms were then held constant for the remaining series at , $\beta_3=$ 4.66(1)*10$^{-4}$ J K$^{-4}$ mol$^{-1}$ and $\beta_5=$4.45(1)*10$^{-7}$ J K$^{-6}$ mol$^{-1}$ respectively. A two level Schottky anomaly, $A_{LT} f(\Delta_{LT},T)$, where $A_{LT}$ is the scaling factor which determines the peak intensity and $\Delta_{LT}$ the size of the gap, accounts for the contribution from defect spins from interlayer Cu$^{2+}$. A second two level Schottky anomaly,  $A_{HT}f(\Delta_{HT},T)$, describes the high temperature features. To reduce the number of independent parameters, the phonon contributions were held fixed across all refinements, as the inserted lithium should result in high frequency modes with only small perturbations of the low temperature phonon spectrum. Further, in initial fits, the magnitude of the gap, $\Delta_{HT}$ = 73 K ($T_m$ = 31 K), of the high temperature Schottky anomaly was found to not vary significantly and thus held constant. Results from the final refinements are given in the SI. While we caution against over-interpretation of many of the obtained values, the magnitude of the high temperature Schottky anomaly, $A_{HT}$, is robust; this was checked by comparing the predictions from fits up to $T$ = 20 K, to the data extending up to $T$ = 300 K in Fig.\ref{HeatCap2}(a).  Upon doping, $A_{HT}$ (Fig.\ref{Figure Combo}(b)) sharply increases then begins to gradually decrease.

This model also fits to the field dependent heat capacity, shown in Fig.\ref{HeatCap2}(b). Similar to the zero field data, the phonon terms, $\beta_3 T^3$ and $\beta_5 T^5$, were calculated based on the field fit to the parent and held constant at the above values for the remaining series. The parameters $\gamma$, $A_{HT}$, and $\Delta_{HT}$ were shared across fields for each sample and each sample was refined independently until convergence. These constraints yielded results consistent with the zero field fits. All the fits clearly demonstrate the field dependence of the low temperature feature which is consistent with a contribution from the magnetic interlayer Cu$^{2+}$. The low temperature magnetization measurements, sensitive to the interlayer Cu on the Zn site, indicate that these interlayer Cu atoms are also systematically reduced as a function of doping.  If these Cu impurities give rise to the finite $\gamma$, it is expected that $\gamma$ would also be reduced with doping as observed. Alternately, if the $\gamma T$ term describes the spin liquid contribution to the heat capacity, a systematic decrease in this value could be explained by the reduction of the spin liquid nature of the material as electrons are introduced into the system. More interestingly, the high temperature Schottky anomaly shows no field dependence and reproduces the trend seen in the zero field data. Direct assignment of the heat capacity terms to specific origins is future work, but it is promising that a single model recapitulates data across temperatures, fields, and composition.

This experimental data is in good agreement with two models for singlet trapping as a function of doping; a Monte Carlo simulation of the trapping of neighboring singlets by Cu$^{1+}$ defects (blue dashed line Fig.\ref{Figure Combo}(b)) and a calculation of singlet trapping by localized electrons on Cu triangles in the kagom\'{e} lattice (black dotted line) (see SI). Since the magnitude of the gap is on the same order as the expected singlet-triplet gap energy in isolated valence bonds in Herbertsmithite\cite{Singh2008}, it is alluring to interpret the growth in high temperature specific heat as arising due to the trapping of valence bonds into a glass or solid-like state. However, further work is needed to exclude other possibilities, such as a localized oscillator mode arising from the inserted Li ions. The singlet trapping models are also in agreement with the magnetization data. Every intercalated Li atom reduces one Cu atom, removing its spin contribution and yielding a one-to-one relationship. So upon doping, the Curie constant will linearly go to zero, in agreement with the experimental data.

In conclusion, we have successfully introduced electrons into the prototypical kagom\'{e} quantum spin liquid Herbertsmithite. Despite the predictions, the doping of this system did not lead to metallicity or superconductivity down to $T$ = 1.8 K. The magnetic field, temperature, and composition dependent specific heat all fit remarkably well to a single model. What are the precise physical origins responsible for this behavior? It is plausible that the location of the inserted Li ions provides a sufficiently strong disorder potential that Anderson localization is never overcome, irrespective of electron count, but other explanations cannot be ruled out\cite{Anderson1958}\cite{Sheckelton2015}. The interesting physics is the following: why does charge doping this spin liquid not change it into a metal? The lower connectivity, with the 2-D kagom\'{e} lattice connects to four magnetic neighbors ($n$ = 4) as compared to six magnetic neighbors of a 2-D triangular lattice ($n$ = 6), may also play a role in the doped series’ behavior. Previous pressure and doping studies on higher connectivity frustrated geometries, such as organic triangular lattice $\kappa$-(ET)$_{2}$Cu$_{2}$(CN)$_{3}$\cite{Kurosaki2005}, Na$_{x}$CoO$_{2}$\cite{Lee2006}, and Na$_4$Ir$_3$O$_8$\cite{Okamoto2007,Podolsky2011} display metallicity. However, to our knowledge, no one has successfully induced metallic behavior in lower connectivity magnetically frustrated structures such as the kagom\'{e} ($n$ = 4) or honeycomb lattice ($n$ = 3). And finally, what is the nature of the ground state of doped Herbertsmithite? Our results demonstrate the need for an improved approach to describe and predict how electron doping effects magnetically frustrated narrow band insulators and implies that, if metallicity is to be induced, the doping method must involve chemical changes far from the kagom\'{e} layers.

\begin{acknowledgments}
\section{Acknowledgments}
This work was supported by the NSF, Division of Materials Research (DMR), Solid State Chemistry (SSMC), CAREER grant under Award DMR-1253562 and the David and Lucile Packard Foundation. TMM was supported by the Institute for Quantum Matter, under Grant No. DE-FG02- 08ER46544. Z.A.K. acknowledges the assistance of A. Huq and K. Page in collecting powder neutron data from POWGEN and NOMAD/SNS. Z.A.K and T.M.M. also thank H.D. Fairbrother and the Surface Analysis Laboratory in the Department of Material Science and Engineering at The Johns Hopkins University, and O. Tchernyshyov for useful discussions.
\end{acknowledgments}

\section{APPENDIX A: MATERIALS AND METHODS}
Phase-pure Herbertsmithite was synthesized hydrothermally in a sealed 21 mL acid digestion vessel. Stoichiometric amounts of ZnCl$_2$ and Cu$_2$(OH)$_2$CO$_3$ in 10 mL of H$_2$O were ramped to $210\,^{\circ}{\rm C}$ at $60\,^{\circ}{\rm C}$ h$^{-1}$, held for 24 h, and cooled to room temp at $6\,^{\circ}{\rm C}$ h$^{-1}$. Several batches were made by this process and thoroughly mixed in order to achieve a large supply of the parent material. All further chemical manipulations were done in Schlenk flasks using air-free techniques. Various amounts of Li were added to a 0.20 M benzophenone (Ph$_2$CO) in THF solution and allowed to stir overnight until all Li dissolved yielding a deep blue or purple solution depending on Li content. The parent Herbertsmithite was then added under the following conditions for the following samples: sample A ($x = 1.8$) was made by intercalation using 1 g of parent material in 50 mL with a molar ratio of 1:1.25 Ph$_2$CO:Li metal and refluxed for 24 h, sample B ($x = 0.2$) was made by intercalation using 1 g of parent material in 30 mL with a molar ratio of 1.1:1 Ph$_2$CO:Li metal and heated at $45\,^{\circ}\mathrm{C}$ for 24 h, sample N ($x = 1.8$) was made by intercalation using 3 g of parent material in 200 mL with a molar ratio of 1:1.25 Ph$_2$CO:Li metal and refluxed for 48 h. 75 mg of sample A ($x = 1.8$) was deintercalated with 15.0 mL and 25.0 mL of 4.50(2) mM I$_2$ in acetonitrile at room temperature until solution became clear to create $x = 1.2$ (A2) and $x = 0.8$ (A3) samples respectively.

Laboratory X-ray powder diffraction patterns were collected using Cu K$\alpha$ radiation ($\lambda$ = 1.5418 $\AA$) on a Bruker D8 Focus diffractometer with a LynxEye detector. Powder neutron diffraction data of sample N at 300 K were collected at the Spallation Neutron Source NOMAD diffractometer (BL-1B) at the Oak Ridge National Laboratory and analyzed with the Rietveld method using GSAS/EXPGUI\cite{Larson2004,Larson2001}. Compositions of the maximally doped specimens were fixed at the values obtained from NPD; the composition of deintercalated samples was determined by the known quantity of oxidant consumed. All other compositions were estimated based on magnetization data.

X-ray photoelectron spectra were collected using Mg K$\alpha$ radiation (1253.6 eV, 15 kV, 300 W) with a pass energy of 58.7 eV, 0.125 eV/step at 50 ms/step on a PHI 5600 XPS. Select samples were ion sputtered with 4 keV Ar$^+$ for 5, 15, 60 and 100 min (ion sputter area 6 x 6 mm$^2$, target current 1.0(3) $\mu$A) with a differential ion gun. XP spectra were energy adjusted to ion sputter cleaned copper metal Cu2p$_{3/2}$ with CasaXPS software. Error in peak position for the Cu 2p and Cu L$_3$M$_{4,5}$M$_{4,5}$ envelopes were estimated to be $\pm$ 3 step sizes (0.375 eV).

Magnetization and heat capacity measurements were measured on powders and cold pressed pellets respectively in a Quantum Design Physical Properties Measurement System. Magnetizations were measured from $T$ = 1.8-300 K under a field of $\mu_0H$ = 0.1 T and susceptibility estimated as $\chi = M/H$. Heat capacity was measured in triplicate at each point using the semi-adiabatic pulse technique. Data was collected from $T$ = 1.8-300 K under $\mu_0H$ = 0 T and from $T$ = 1.8-20 K under $\mu_0H$ = 1, 5, 9 T. Two probe contact resistivity measurements with a voltmeter on the series of cold pressed polycrystalline samples at room temperature indicated a resistance of $>$2M$\Omega$.

\section{}

\bibliography{ElectronDoping.v2}

\end{document}